# A Four-Unit-Cell Periodic Pattern of Quasiparticle States Surrounding Vortex Cores in $Bi_2Sr_2CaCu_2O_{8+d}$

J. E. Hoffman[1], E. W. Hudson[1,3*], K. M. Lang[1], V. Madhavan[1],
H. Eisaki[2‡], S. Uchida[2] & J.C. Davis[1,3§]

[1] *Department of Physics, University of California, Berkeley, CA 94720-7300, USA.*
[2] *Department of Superconductivity, University of Tokyo, Tokyo, 113-8656 Japan.*
[3] *Materials Sciences Division, Lawrence Berkeley National Laboratory. Berkeley, CA 94720, USA.*
[*] *Present address: Department of Physics, MIT, Cambridge, MA 02139-4307, USA.*
[‡] *Present address: Department of Applied Physics, Stanford University, Stanford, CA 94305-4060, USA.*
[§] *To whom correspondence should be addressed.*
*E-mail: jcdavis@socrates.berkeley.edu*

**Scanning tunneling microscopy is used to image the additional quasiparticle states generated by quantized vortices in the high-$T_c$ superconductor $Bi_2Sr_2CaCu_2O_{8+d}$. They exhibit a Cu-O bond oriented 'checkerboard' pattern, with four unit cell ($4a_0$) periodicity and a ~30Å decay length. These electronic modulations may be related to the magnetic field-induced, $8a_0$ periodic, spin density modulations of decay length ~70Å recently discovered in $La_{1.84}Sr_{0.16}CuO_4$. The proposed explanation is a spin density wave localized surrounding each vortex core. General theoretical principles predict that, in the cuprates, a localized spin modulation of wavelength l should be associated with a corresponding electronic modulation of wavelength l/2, in good agreement with our observations.**

Theory indicates that the electronic structure of the cuprates is susceptible to transitions into a variety of ordered states *(1-10)*. Experimentally, antiferromagnetism (AF) and high temperature superconductivity (HTSC) occupy well known regions of the phase diagram but, outside these regions, several unidentified ordered states exist. For example, at low hole densities and above the superconducting transition temperature, the unidentified "pseudogap" state exhibits gapped electronic excitations *(11)*. Other unidentified ordered states, both insulating *(12)* and conducting *(13)*, exist in magnetic fields sufficient to quench superconductivity. Categorization of the cuprate electronic ordered states and clarification of their relationship to HTSC are among the key challenges in condensed matter physics today.

Because the suppression of superconductivity inside a vortex core can allow one of the alternative ordered states *(1-10)* to appear there, the electronic structure of HTSC vortices has attracted wide attention. Initially, theoretical efforts focused on the quantized vortex in a Bardeen-Cooper-Schrieffer (BCS) superconductor with $d_{x^2-y^2}$ symmetry *(14-18)*. These models included predictions that, because of the gap-nodes, the local density of electronic states (LDOS) inside the core is strongly peaked at the Fermi level. This peak, which would appear in tunneling studies as a zero bias conductance peak (ZBCP), should display a four-fold symmetric "star shape" oriented toward the gap nodes and decaying as a power law with distance.

Scanning tunneling microscopy (STM) studies of HTSC vortices have revealed a very different electronic structure from that predicted by the pure d-wave BCS models. Vortices in $YBa_2Cu_3O_7$ (YBCO) lack ZBCPs but exhibit additional quasiparticle states at ±5.5meV *(19)*, whereas those in $Bi_2Sr_2CaCu_2O_{8+\delta}$ (Bi-2212) also lack ZBCPs *(20)*. More recently, the additional quasiparticle states at Bi-2212 vortices were discovered at energies near ±7meV *(21)*. Thus, a common phenomenology for low energy quasiparticles associated with vortices is becoming apparent. Its features include: (a) the absence of ZBCP's, (b) a radius for the actual vortex core (where the coherence peaks are absent) of ~10Å *(21)*, (c) low energy quasiparticle states at ±5.5meV (YBCO) and ±7meV (Bi-2212), (d) a radius of up to ~75Å within which these states are detected *(21)*, and (e) the absence of a four-fold symmetric star-shaped LDOS.

Because d-wave BCS models do not explain this phenomenology, new theoretical approaches have been developed. Zhang *(5)* and Arovas *et al. (22)* first focused attention on magnetic phenomena associated with HTSC vortices with proposals that a magnetic field induces antiferromagnetic order localized by the core. More generally, new theories describe vortex-induced electronic and magnetic phenomena when the anticipated effects of strong correlations and strong antiferromagnetic spin fluctuations are included *(22-26)*. Common elements of their predictions include: (a) the proximity of a phase transition into a magnetic ordered state can be revealed when the superconductivity is weakened by the influence of a vortex *(22-26)*, (b) the resulting magnetic order, either spin *(22, 23, 25)* or orbital *(24, 26)*, will coexist with superconductivity in some region near the core, and (c) this localized magnetic order will generate associated spatial modulations in the quasiparticle density of states *(23-26)*. Given the relevance of such predictions to the identification of alternative ordered states, determination of the magnetic and electronic structure of the HTSC vortex is an urgent priority.

Information on the magnetic structure of HTSC vortices has recently become available from neutron scattering and nuclear magnetic resonance (NMR) studies. Near optimum doping, some cuprates show strong inelastic neutron scattering (INS) peaks at the four **k**-space points (1/2±δ, 1/2) and (1/2, 1/2±δ), where δ~1/8 and **k**-space distances are measured in units $2\pi/a_0$. This demonstrates the existence, in real space, of fluctuating magnetization density with spatial periodicity of $8a_0$ oriented along the Cu-O bond directions, in the superconducting phase. The first evidence for field-induced magnetic order in the cuprates came from INS experiments on $La_{1.84}Sr_{0.16}CuO_4$ by Lake *et al. (27)*. When $La_{1.84}Sr_{0.16}CuO_4$ is cooled into the superconducting state, the scattering intensity at these characteristic **k**-space locations disappears at energies below ~7meV, opening up a "spin gap". Application of a 7.5T magnetic field below 10K causes the scattering intensity to reappear with strength almost equal to that in the normal state. These field-induced spin fluctuations have a spatial periodicity of $8a_0$ and wavevector pointing along the Cu-O bond direction. Their magnetic coherence length $L_M$ is at least $20a_0$ although the vortex core diameter is only $\sim 5a_0$. More recently, studies by Khaykovich *et al. (28)* on a related material, $La_2CuO_{4+y}$, found field-induced enhancement of elastic neutron scattering (ENS) intensity at these same incommensurate **k**-space locations, but with $L_M > 100a_0$. Thus, field-induced static

AF order with $8a_0$ periodicity exists in this material. Finally, NMR studies by Mitrovic *et al.* *(29)* explored the spatial distribution of magnetic fluctuations near the core. NMR is used because $1/T_1$, the inverse spin-lattice relaxation time, is a measure of spin-fluctuations, and the Larmor frequency of the probe nucleus is a measure of their locations relative to the vortex center. In YBCO at B=13T, the $1/T_1$ of $^{17}O$ rises rapidly as the core is approached, then diminishes inside the core. These experiments are all consistent with vortex-induced spin fluctuations occurring outside the core.

Theoretical attention was first focused on the regions outside the core by a phenomenological model that proposed that the circulating supercurrents weaken the superconducting order parameter and allow the local appearance of a coexisting spin density wave (SDW) and HTSC phase *(23)* surrounding the core. In a more recent model, which is an extension of (*5, 22*), the effective mass associated with spin fluctuations results in an AF localization length that might be substantially greater than the core radius *(30)*. An associated appearance of charge density wave order was also predicted *(31)* whose effects on the HTSC quasiparticles should be detectable in the regions surrounding the vortex core (*23*).

To test these ideas, we apply our recently developed techniques of low-energy quasiparticle imaging at HTSC vortices *(21)*. We choose to study Bi-2212, because YBCO and LSCO have proven non-ideal for spectroscopic studies because their cleaved surfaces often exhibit non-superconducting spectra. Our "as-grown" Bi-2212 crystals are generated by the floating zone method, are slightly overdoped with $T_c$=89K, and contain 0.5% of Ni impurity atoms. They are cleaved (at the BiO plane) in cryogenic ultra-high vacuum below 30K, and immediately inserted into the STM head. Fig. 1A shows a topographic image of the 560Å square area where all the STM measurements reported here were carried out. The atomic resolution and the supermodulation (with wavelength ~26Å oriented at 45° to the Cu-O bond directions) are evident throughout.

To study effects of the magnetic field $B$ on the superconducting electronic structure, we first acquire zero-field maps of the differential tunneling conductance ($G=dI/dV$) measured at all locations ($x$, $y$) in the field of view (FOV) of Fig. 1A. Because $LDOS(E=eV)\propto G(V)$ where $V$ is the sample bias voltage, this results in a two dimensional map of the local density of states $LDOS(E,x,y,B=0)$. We acquire these LDOS maps at energies ranging from –12meV to +12meV in 1meV increments. The $B$ field is then ramped to its target value and, after any drift has stabilized, we re-measure the topograph with the same resolution. The FOV where the high-field LDOS measurements are to be made is then matched to that in Fig. 1A within 1Å (~$0.25a_0$) by comparing characteristic topographic/spectroscopic features. Finally we acquire the high-field LDOS maps, $LDOS(E,x,y,B)$, at the same series of energies as the zero-field case.

To focus preferentially on $B$ field effects, we define a new type of two dimensional map:

$$S_{E_1}^{E_2}(x,y,B)=\sum_{E_1}^{E_2}\left[LDOS(E,x,y,B)-LDOS(E,x,y,0)\right]dE \qquad (1)$$

which represents the integral of all additional spectral density induced by the $B$ field between the energies $E_1$ and $E_2$ at each location ($x$, $y$). We use this technique of combined electronic background subtraction and energy integration to enhance the signal-to-noise ratio of the vortex-induced states. In Bi-2212, these states are broadly distributed in energy around ±7meV *(21)*, so $S_{\pm 1}^{\pm 12}(x, y, B)$ effectively maps the additional spectral strength under their peaks.

Fig. 1B is an image of $S_1^{12}(x, y, 5)$ measured in the FOV of Fig. 1A. The locations of seven vortices are evident as the darker regions of dimension ~100Å. Each vortex displays a spatial structure in the integrated LDOS consisting of a "checkerboard" pattern oriented along Cu-O bonds. We have observed spatial structure with the same periodicity and orientation, in the vortex-induced LDOS on multiple samples and at fields ranging from 2 to 7 Tesla. In all 35 vortices studied in detail, this spatial and energetic structure exists, but the "checkerboard" is more clearly resolved by the positive-bias peak.

We show the power spectrum from the two-dimensional Fourier transform of $S_1^{12}(x, y, 5)$, $PS[S_1^{12}(x, y, 5)] = [FT(S_1^{12}(x, y, 5))]^2$, in Fig. 2A and a labeled schematic of these results in Fig. 2B. In these **k**-space images, the atomic periodicity is detected at the points labeled by A, which by definition are at (0,±1) and (±1,0). The harmonics of the supermodulation are identified by the symbols $B_1$ and $B_2$. These features (A, $B_1$, and $B_2$) are observed in the Fourier transforms of all LDOS maps, independent of magnetic field, and they remain as a small background signal in $PS[S_1^{12}(x, y, 5)]$ because the zero-field and high-field LDOS images can only be matched to within 1Å before subtraction. Most importantly, $PS[S_1^{12}(x, y, 5)]$ reveals new peaks at the four **k**-space points which correspond to the spatial structure of the vortex-induced quasiparticle states. We label their locations C. No similar peaks in the spectral weight exist at these points in the two-dimensional Fourier transform of these zero-field LDOS maps.

To quantify these results, we fit a Lorentzian to $PS[S_1^{12}(x, y, 5)]$ at each of the four points labeled C in Fig. 2B. We find that they occur at **k**-space radius 0.062Å$^{-1}$ with width $\sigma$=0.011±0.002Å$^{-1}$. Fig. 2C shows the value of $PS[S_1^{12}(x, y, 5)]$ measured along the dashed line in Fig. 2B. The central peak associated with long wavelength structure, the peak associated with the atoms, and the peak due to the vortex-induced quasiparticle states are all evident. The vortex-induced states identified by this means occur at (±1/4,0) and (0,±1/4) to within the accuracy of the measurement. Equivalently, the "checkerboard" pattern evident in the LDOS has spatial periodicity $4a_0$ oriented along the Cu-O bonds. Furthermore, the width $\sigma$ of the Lorentzian yields a spatial correlation length for these LDOS oscillations of $L=(1/\pi\sigma) \approx 30\pm5$Å (or $L \approx 7.8\pm1.3a_0$). This is substantially greater than the measured *(21)* core radius. It is also evident in Figs. 1B and 2A that the LDOS oscillations have stronger spectral weight in one Cu-O direction than in the other. The ratio of amplitudes of $PS[S_1^{12}(x, y, 5)]$ between (±1/4,0) and (0,±1/4) is approximately 3.

How might these observations relate to the spin structure *(27-29)* of the HTSC vortex? The original suggestion of an AF insulating region inside the core *(5, 22)* cannot be tested directly by our techniques, although the Fermi-level LDOS measured there is low *(20,*

*21)*. A more recent proposal is that when the HTSC order parameter near a vortex is weakened by circulating superflow, a coexisting SDW+HTSC phase appears resulting in a local magnetic state **M**(r) surrounding the core *(23)*. A second proposal is that the periodicity, orientation, and spatial extent of the vortex-induced **M**(r) are determined by the dispersion and wavevector of the pre-existing zero field AF fluctuations *(30)*. In both cases, the $8a_0$ spatial periodicity of **M**(r) is not fully understood but is consistent with models of evolution of coupled spin and charge modulations in a doped antiferromagnetic Mott insulator *(8, 32)*. A final possibility is that $8a_0$ periodic "stripes" *(2, 3, 6, 7)* are localized surrounding the core, but that two orthogonal configurations are apparent in the STM images because of fluctuations in a nematic stripe phase *(33, 34)*, or because of bilayer effects. Of central relevance to the results reported here is the fact that, in all of these models, the magnetic state bound to the vortex has $8a_o$ periodicity and is oriented parallel to both the Cu-O directions.

Fig 3A shows a schematic of the superflow field and the magnetization **M**(r) localized at the vortex. Almost all microscopic models predict that magnetic order localized near a vortex will create characteristic perturbations to the quasiparticle LDOS (23-26, 31). In addition, general principles about coupled charge- and spin-density-wave order parameters *(2, 3, 6-8, 32)* indicate that spatial variations in **M**(r) must have double the wavelength of any associated variations in the LDOS(r). Thus, the perturbations to the LDOS(r) near a vortex should have $4a_o$ periodicity and the same orientation and spatial extent as **M**(r) as represented schematically in Fig 3A. In an LDOS image this would become apparent as a checkerboard pattern (Fig. 3B). In Fig. 3C, we show the autocorrelation of a region of Fig. 1B that contains one vortex, to display the spatial structure of the Bi-2212 vortex-induced LDOS. It is in good agreement with the quasiparticle response described by Fig. 3, A and B. Therefore, assuming equivalent vortex phenomena in LSCO, YBCO and Bi-2212, the combined results from INS, ENS, NMR, and STM lead to an internally consistent new picture for the electronic and magnetic structure of the HTSC vortex

Independent of models of the vortex structure, the data reported here are important for several reasons. First, the $4a_0$ periodicity and register to the Cu-O bond directions of the vortex-induced LDOS are likely signatures of strong electronic correlations in the underlying lattice. Such a $4a_o$ periodicity in the electronic structure is a frequent prediction of coupled spin-charge order theories for the cuprates *(2, 3, 6-8, 32-34)*, but has not been previously observed in the quasiparticle spectrum of any HTSC system. Second, some degree of one-dimensionality is evident in these incommensurate LDOS modulations because one Cu-O direction has stronger spectral intensity than the other. Finally, the vortex-induced LDOS is detected by STM at least 50 Å away from the core. This means that, at only 5 T, ~25% of the sample is under the influence of whatever phenomenon generates the "checkerboard" of LDOS modulations, even though the vortex cores themselves make up only ~2% of its area.

35. We acknowledge and thank S. Sachdev, S. Kivelson, D.-H. Lee, E. Demler, S.-C. Zhang, J. Sethna, G. Aeppli, J. Orenstein, W. Halperin, and S. H. Pan for helpful conversations and communications. This work was funded by the Office of Naval Research, the Materials Sciences Division of Lawrence Berkeley National Laboratory, the CULAR Program of Los Alamos National Laboratory, a Grant-in-Aids for Scientific Research, a COE Grant from the Ministry of Education, as well as an International Joint Research Grant from the New Energy and Industrial Technology Organization in Japan. JEH acknowledges support by a Hertz Fellowship, KML by an IBM Fellowship, and JCD by a Miller Professorship.

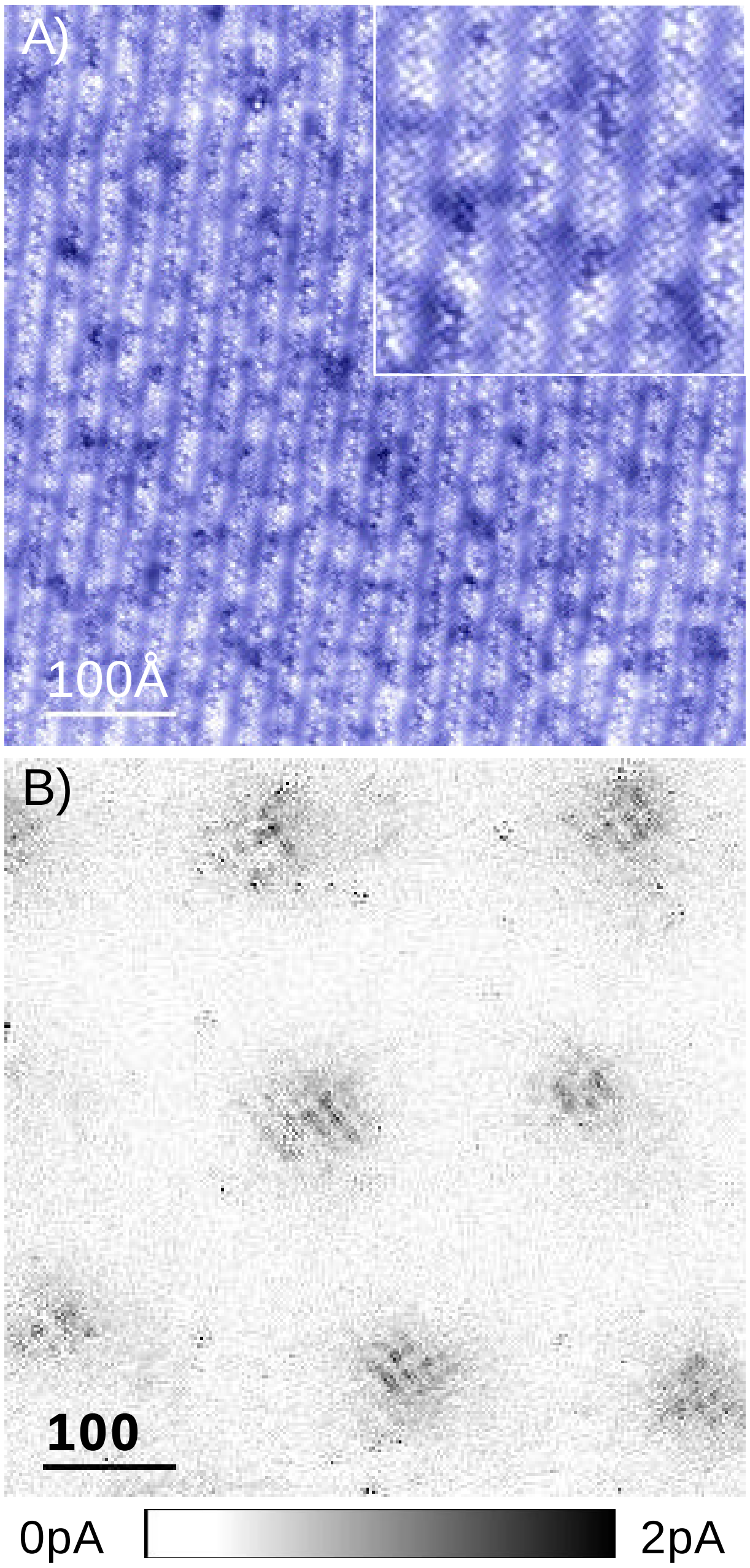

**Figure 1:** Topographic and spectroscopic images of the same area of a Bi-2212 surface.

**(A)** A topographic image of the 560Å field of view (FOV) in which the vortex studies were carried out. The supermodulation can be seen clearly along with some effects of electronic inhomogeneity. The Cu-O-Cu bonds are oriented at 45° to the supermodulation. Atomic resolution is evident throughout, and the inset shows a 140Å square FOV at 2x magnification to make this easier to see. The mean Bi-Bi distance apparent here is $a_0$=3.83Å, and is identical to the mean Cu-Cu distance in the CuO plane ~5Å below.

**(B)** A map of $S_1^{12}(x,y,5)$ showing the additional LDOS induced by the seven vortices. Each vortex is apparent as a 'checkerboard' at 45° to the page orientation. Not all are identical, most likely due to the effects of electronic inhomogeneity. The units of $S_1^{12}(x,y,5)$ are picoamps because it represents $\Sigma$dI/dV ·ΔV. In this energy range, the maximum integrated LDOS at a vortex is ~3pA, as compared with the zero field integrated LDOS of ~1pA. The latter is subtracted from the former to give a maximum contrast of ~2pA. We also note that the integrated differential conductance between 0meV and -200meV is 200pA because all measurements reported in this paper were obtained at a junction resistance of 1GΩ set at bias voltage –200mV.

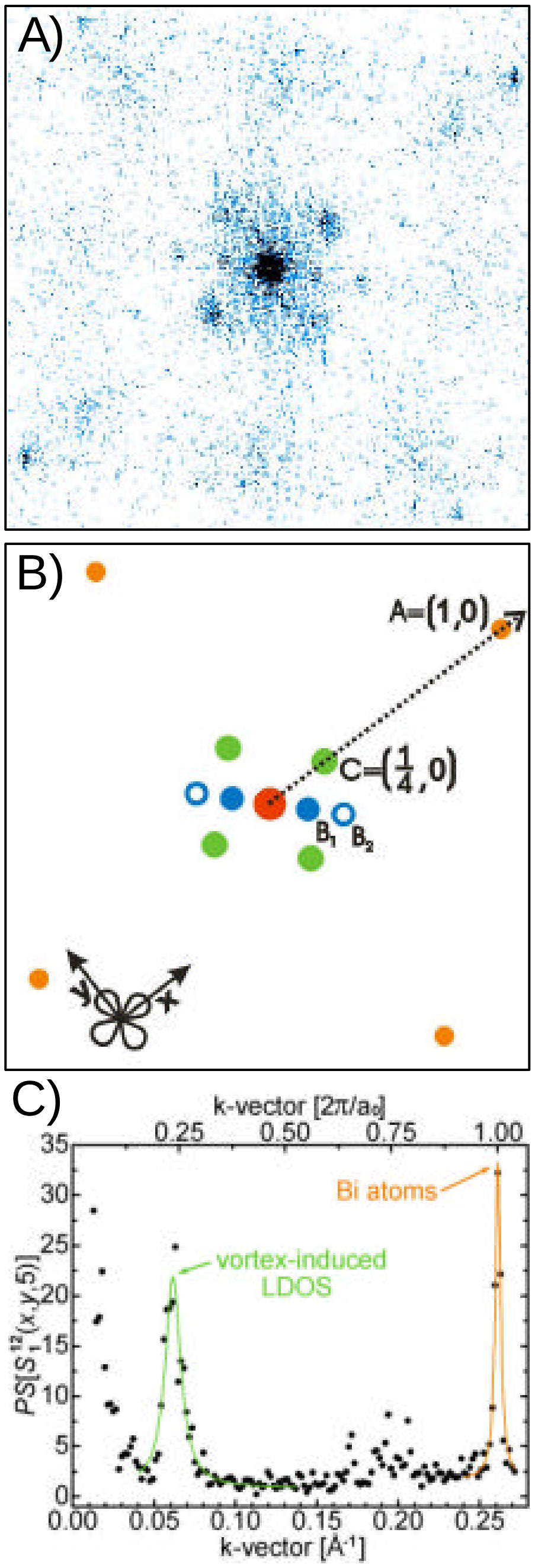

**Figure 2:** Fourier transform analysis of vortex-induced LDOS.

**(A)** $PS[\, S_1^{12}(x, y,5)\, ]$, the two dimensional power spectrum of the $S_1^{12}(5, x, y)$ map shown in Fig. 1B. The four points near the edges of the figure are the k-space locations of the square Bi lattice. The vortex effects surround the **k**=0 point at the center of the figure.

**(B)** A schematic of the $PS[\, S_1^{12}(x, y,5)\, ]$ shown in Fig. 2A. Distances are measured in units of $2\pi/a_0$. Peaks due to the atoms at (0,±1) and (±1,0) are labeled A. Peaks due to the supermodulation are observed at $B_1$ and $B_2$. The four peaks at C occur only in a magnetic field and represent the vortex-induced effects at **k**-space locations (0,±1/4) and (±1/4,0).

**(C)** A trace of $PS[\, S_1^{12}(x, y,5)\, ]$ along the dashed line in Fig. 2B. The strength of the peak due to vortex-induced states is demonstrated, as is its location in the **k**-space unit cell relative to the atomic locations. The spectrum along the line toward (0,1) is equivalent but there is less spectral weight in the peak in $PS[\, S_1^{12}(x, y,5)\, ]$ at (0,1/4).

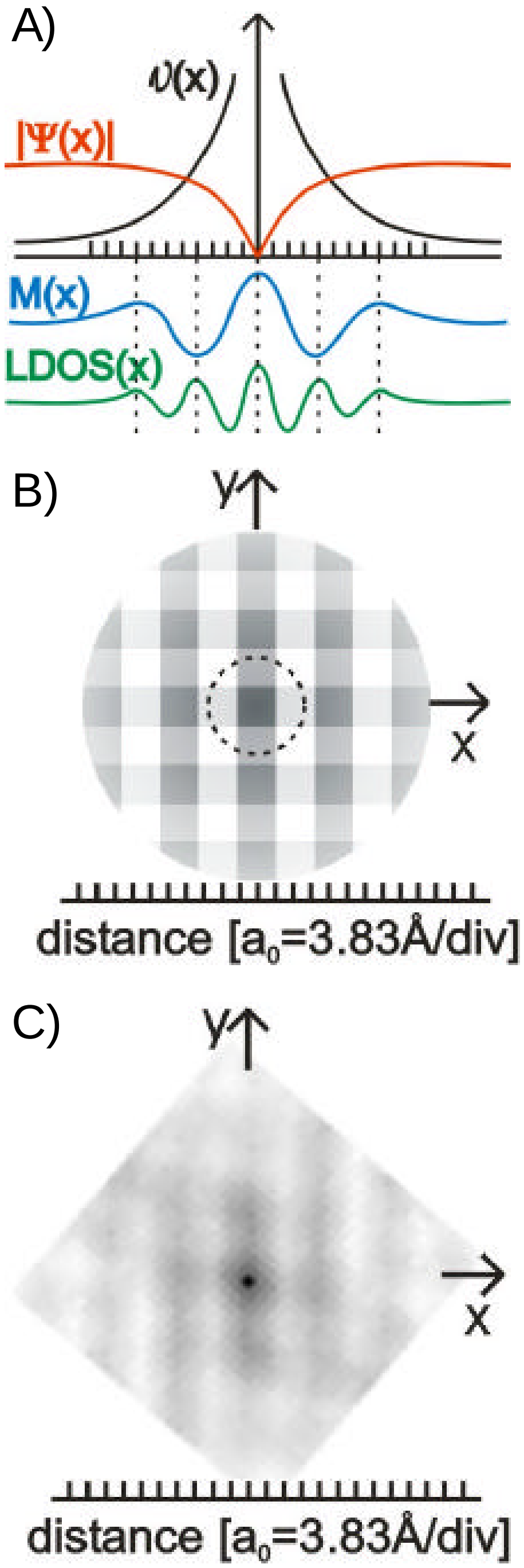

**Figure 3:** A schematic model of the electronic/magnetic structure of the HTSC vortex.

**(A)** Superfluid velocity $n$(x) rises and the HTSC order parameter $|\Psi(x)|$ falls as the core is approached. The periodicity of the spin density modulation deduced from (*27*) is shown schematically as $\mathbf{M}$(x). The anticipated periodicity of the LDOS modulation due to such an $\mathbf{M}$(x) is shown schematically as LDOS(x).

**(B)** A schematic of the two-dimensional "checkerboard" of LDOS modulations that would exist at a circularly symmetric vortex core with an $8a_0$ spin modulation as modeled in Fig 3A. The dashed line shows the location of the ~$5a_0$ diameter vortex core. The dark regions represent higher intensity low energy LDOS due to the presence of a vortex. They are $2a_0$ wide and separated by $4a_0$.

**(C)** The two dimensional autocorrelation of a region of $S_1^{12}(x, y, 5)$ that contains one vortex. Its dimensions are scaled to match the scale of Fig. 3A, and it is rotated relative to Fig. 1 so that the Cu-O bond directions are here horizontal and vertical.